\documentclass[oldversion]{aa}
\usepackage{epsfig}
\usepackage{natbib}
\usepackage{lscape}
\usepackage{amssymb}
\usepackage{epstopdf}
\usepackage[caption=false]{subfig}
\usepackage[caption=false]{subfig}
\usepackage{float}

\usepackage{longtable}
\usepackage{color}
\usepackage{fancyheadings}
\bibpunct{(}{)}{;}{a}{}{,}

\begin{document}

\title{Understanding stellar activity-induced radial velocity jitter using simultaneous \textit{K2} photometry and HARPS RV measurements}

\author{M. Oshagh\inst{1,2}, N. C. Santos\inst{2,3}, P. Figueira\inst{2}, S. C. C. Barros\inst{2}, J.-F. Donati\inst{4,5}, V. Adibekyan\inst{2}, J. P. Faria\inst{2,3}, C. A. Watson\inst{6}, H. M. Cegla\inst{6,7,8}, X. Dumusque\inst{7}, E. H\'ebrard\inst{9}, O. Demangeon\inst{2},  S. Dreizler\inst{1}, I. Boisse\inst{10}, M. Deleuil\inst{10}, X. Bonfils\inst{11}, F. Pepe\inst{7}, S. Udry \inst{7}}

\institute{
Institut f\"ur Astrophysik, Georg-August-Universit\"at,
Friedrich-Hund-Platz 1, 37077 G\"ottingen, Germany
\and
Instituto de Astrof\' isica e Ci\^encias do Espa\c{c}o, Universidade do Porto, CAUP, Rua das Estrelas, PT4150-762 Porto, Portugal 
\and
Departamento de F{\'i}sica e Astronomia, Faculdade de Ci{\^e}ncias, Universidade do
Porto,Rua do Campo Alegre, 4169-007 Porto, Portugal
\and
Universit\'e de Toulouse, UPS-OMP, IRAP, 14 avenue E.~Belin, Toulouse, F--31400 France 
\and
CNRS, IRAP / UMR 5277, Toulouse, 14 avenue E.~Belin, F--31400 France
\and
Astrophysics Research Centre, School of Mathematics \& Physics, Queen's University Belfast, University Road, Belfast, BT7 1NN, UK
\and
Observatoire de Gen\`{e}ve, Universit\'{e} de Gen\`{e}ve, 51 chemin des Maillettes, 1290 Versoix, Switzerland
\and
Swiss National Science Foundation NCCR-PlanetS CHEOPS Fellow, Switzerland
\and
Department of Physics and Astronomy, York University, Toronto, ON M3J 1P3, Canada
\and
Aix-Marseille Université, CNRS, LAM (Laboratoire d'Astrophysique de Marseille) UMR 7326, 13388, Marseille, France
\and
UJF-Grenoble 1 / CNRS-INSU, Institut de Plan\'etologie et d'Astrophysique de Grenoble (IPAG) UMR 5274, Grenoble, F-38041, France\\
}

\date{Received XXX; accepted XXX}

\abstract {One of the best ways to improve our understanding of the stellar activity-induced signal in radial velocity (RV) measurements is through simultaneous high-precision photometric and RV observations. This is of prime importance to mitigate the RV
signal induced by stellar activity and therefore unveil the presence of low-mass exoplanets. The \textit{K2} Campaign 7 and 8 field-of-views
were located in the southern hemisphere, and provided a unique opportunity to gather unprecedented simultaneous high precision photometric observation with \textit{K2} and high-precision RV measurements with the HARPS spectrograph to study the relationship between photometric variability and RV jitter. We observed nine stars with different levels of activity; from quiet to very active. We first probe the presence of any meaningful relation between measured RV jitter and the simultaneous photometric variation, and also other activity indicators (such as BIS, FWHM, $logR'_{HK}$ and F8), by evaluating the strength and significance of the monotonic correlation between RVs and each indicator. We found that for the case of very active stars, strong and significant correlations exist between almost all the observables and measured RVs; however, when we move towards lower activity levels the correlations become random, and we could not reach any conclusion regarding the tendency of correlations depending on the stellar activity level. Except for the F8 which its strong correlation with RV jitter persists over a wide range of stellar activity level, and thus our result suggests that F8 might be a powerful proxy for activity induced RV jitter over a wide range of stellar activity. Moreover, we examine the capability of two state-of-the-art modeling techniques, namely the \textit{FF'} method and \textit{SOAP2.0}, in accurately predicting the RV jitter amplitude using the simultaneous photometric observation. We found that for the very active stars both techniques can reasonably well predict the amplitude of the RV jitter, however, at lower activity levels the \textit{FF'} method underpredicts the RV jitter amplitude.

 }


\keywords{methods: observational, numerical- planetary system- techniques: photometry, spectroscopy}

\authorrunning{M. Oshagh et al.}
\titlerunning{Simultaneous HARPS and \textit{K2} observations.}
\maketitle
\section{Introduction}
It is well-known that the presence of stellar active regions (such as star spots or plages) on a rotating star can generate astrophysical noise in high-precision photometric and radial velocity (RV) time series. The activity-induced RV jitter can hamper the detection of low-mass planets, complicate the confirmation of transiting planets, and may even mimic a planetary signal \citep[e.g.][]{Queloz-01, Santos-02, Huelamo-08, Figueira-10, Santos-10, Boisse-11, Dumusque-11, Santos-14, Robertson-14, Diaz-16}. In photometric observations the activity noise mayalso cause severe difficulties in accurately characterizing transiting planets through transit light-curves analysis  \citep[e.g.][]{Czesla-09, Oshagh-13b, Oshagh-14, Barros-14, Oshagh-15}.

\begin{table*}
\caption{Main stellar parameters, the EPIC number, the number of \textit{K2} campaign, and number of HARPS RV observations of our target list. The table is  ranked based on the target $logR'_{HK}$ values (decreasing). }              
\label{table:1}      
\centering                                      
\begin{tabular}{c c c c c c c c c c} 
\hline\\
 Name  & EPIC Number & Spec type & $T_{eff} (K)$ & Mass (${M_{\odot}}$) & Radius (${R_{\odot}}$) & Mag (V) & $logR'_{HK}$& \textit{K2} \# FOV & \# of RV\\
\hline\hline   
\\
HD173427 & 214112021 & G1V & 6244& 1.189 & 1.367 & 8.57 & -4.33 & 7 & 7\\
HD181544 & 213410372 & G1V & 6212 &1.218& 1.603&  7.11& -4.61& 7 & 6\\
HD177033 & 213812240 & K2/K3V & 5090& 0.740& 0.709 & 10.07 & -4.66& 7 & 3\\
HD6101 & 220417763 & K2V & 4991& 0.844&0.781& 8.131& -4.76&8& 4\\
HD6480 & 220409971 & F5/7 V & 5970   &1.081& 1.159& 7.25&-4.86& 8& 12\\
HD183877 & 213873758 & G8V & 5748& 1.009& 0.977& 7.15&-4.94& 7 &7\\
HD4628 & 220429217 & K2.5V &5057& 0.787&0.731& 5.74&-4.94& 8 & 5\\
HD4256 & 220260370 & K3V& 5047 & 0.828& 0.760& 8.001&-5.08& 8 & 4\\
\hline                                             
\\
HD179205 & 214776835 & G1/2V &5988 &0.882 &0.977 & 8.59 & -4.55&7& 5\\
\hline                                             
\end{tabular}
\end{table*}

Many studies have suggested the existence of a correlation between the amplitude of photometric variability of stars and the amplitude of RV jitter. For instance, the classic work by \citet{Saar-98} established the first estimate of this correlation, and provided a relation to predict RV jitter for a given star as a function of $v \sin i$, spectral type, and photometric variability. Later, \citet{Boisse-09} and \citet{Lanza-11} performed simultaneous photometric and RV observations of HD189733 and demonstrated that simultaneous photometric time series can deliver a wealth of information about the configuration and distribution of active regions on the surface of a star, and thus would allow us to improve
our understanding of the RV jitter. However, it is important to remember that HD189733 is a very active star and, therefore, the conclusion of these studies could not easily be extrapolated to other stars with different activity levels, and in particular to relatively low-activity stars like our Sun.

To further study the relationship between photometric variability and the expected level of RV jitter, \citet{Cegla-14} utilised high precision light curves from the \textit{Kepler} space telescope. Unfortunately, due to the faintness of stars in the \textit{Kepler} field, direct RV measurements were unavailable for most targets. To counter this, these authors used readily available GALEX UV flux measurements and a series of empirical relationships (between excess UV flux, the calcium activity indicator ($logR'_{HK}$), and RV jitter) to indirectly estimate the RV jitter.

Later, \citet{Bastien-14} went one step further, and collected the California Planet Search archival RV measurements of stars that lay in the \textit{Kepler} field of view. They used high-precision RV measurements obtained at the Keck and Lick observatories and high-precision photometric observations of twelve \textit{Kepler} stars. They searched for a common periodicity between RV jitter and photometric variability of each individual star. The major drawback of their study was that their RV measurements were not taken simultaneously with Kepler's photometric observations. From solar and stellar observations, it is quite well-known that the stellar magnetic activity evolves as a function of time, even over a
single sollar rotation \citep[e.g.][]{Baliunas-95}, and therefore observations
performed during different epochs can be completely independent measurements and may nnot exhibit any meaningful or physical correlations.

A new opportunity was provided by \textit{K2}, which is an extended mission of the original \textit{Kepler} space telescope \citep{Borucki-10} after the failure of \textit{Kepler}'s two reaction wheels. \textit{K2} aims to observe different fields, all in the ecliptic plane, each for the duration of approximately 80 days \citep{Howell-14}. \textit{K2} Campaign 7 and 8 field-of-views were observed between 4th of October and 26th of December 2015, and between 3rd of January and 23rd of March 2016, respectively.  Both \textit{K2} Campaign 7 and 8 had the particularity of being positioned in the southern hemisphere, which provided us the unique opportunity to carry out simultaneous high-precision RV measurements with the HARPS spectrograph mounted on the 3.6m ESO telescope. We were allocated three nights of observation time with HARPS to execute RV measurements simultaneous with \textit{K2} observations of nine stars.

The main goal of the current paper is to identify and characterize the possible correlation between
photometric variability and RV jitter, using simultaneous space-based \textit{K2} high-precision photometry
and HARPS's high-precision RV measurements. Such a full characterization and relationship will be crucial for selecting the best transiting candidates, to be
followed up by RV observations, for upcoming missions such as TESS \citep{Ricker-14} and PLATO 2.0 \citep{Rauer-14}, and
will greatly improve the efficiency of the RV follow-up of planet candidates with the next generation of stabilised spectrographs such as  ESPRESSO \citep{Pepe-14}.  We also explore the variation of different activity indicators, such as direct chromospheric activity indicator ($logR'_{HK}$), and atmospheric line profile diagnostics such as BIS and FWHM.

\begin{figure*}
	\center
	\vspace{-0.5cm}
	\hspace{-1.3cm}  \subfloat{\includegraphics[width=0.36\textwidth, height=90mm]{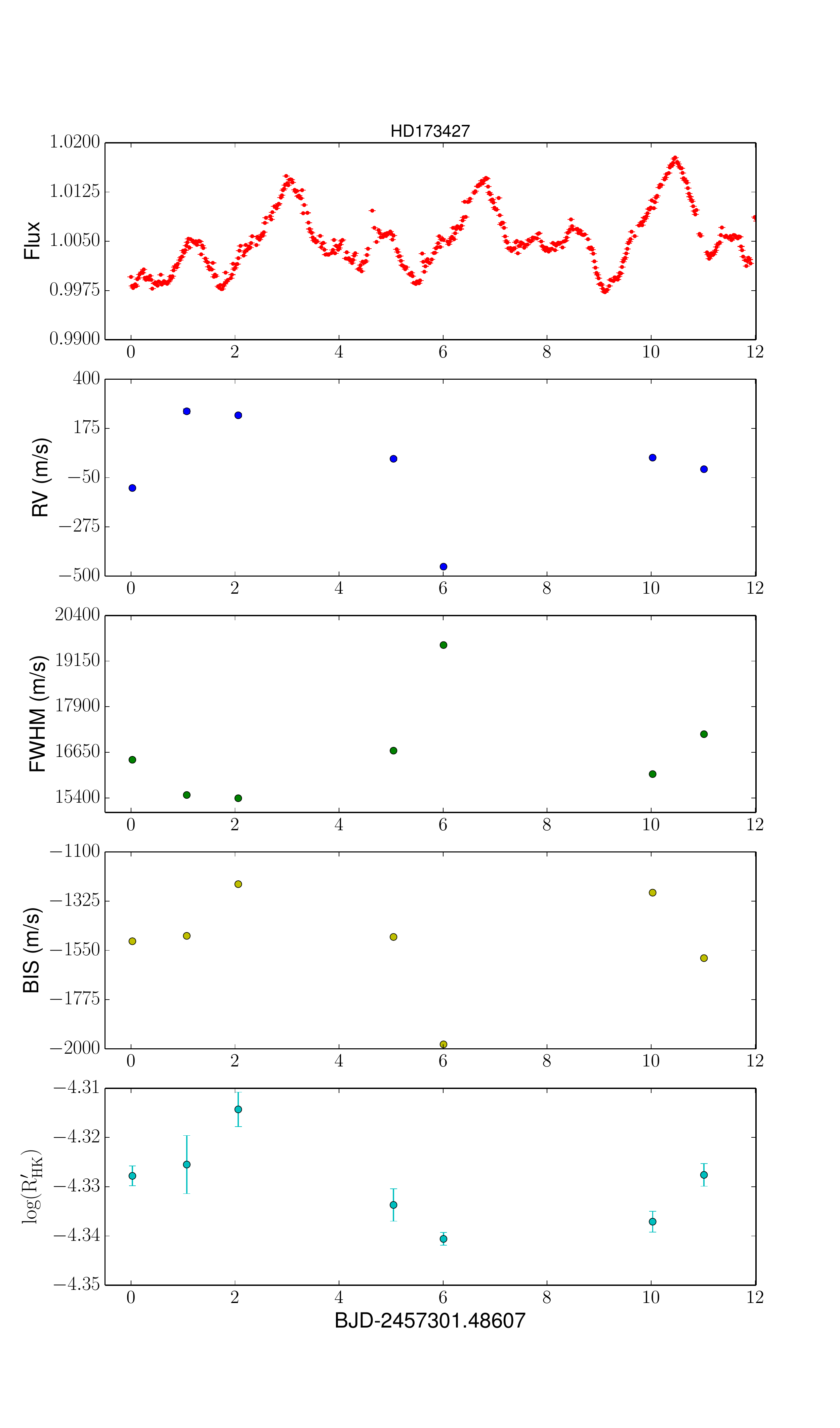}}\hspace{-0.2cm}\medskip
  \subfloat{\includegraphics[width=0.36\textwidth, height=90mm]{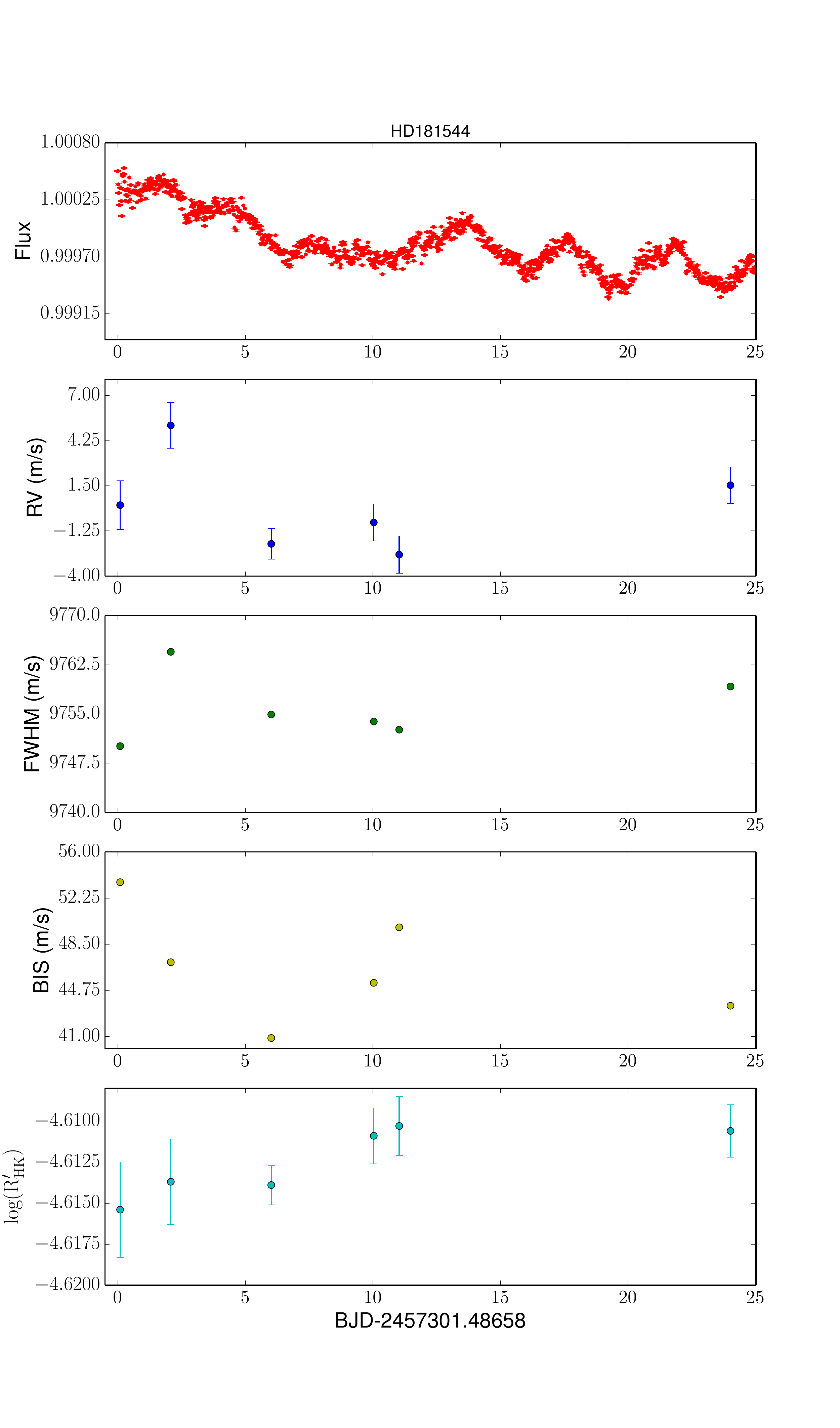}}\medskip\hspace{-0.2cm}\medskip
  \subfloat{\includegraphics[width=0.36\textwidth, height=90mm]{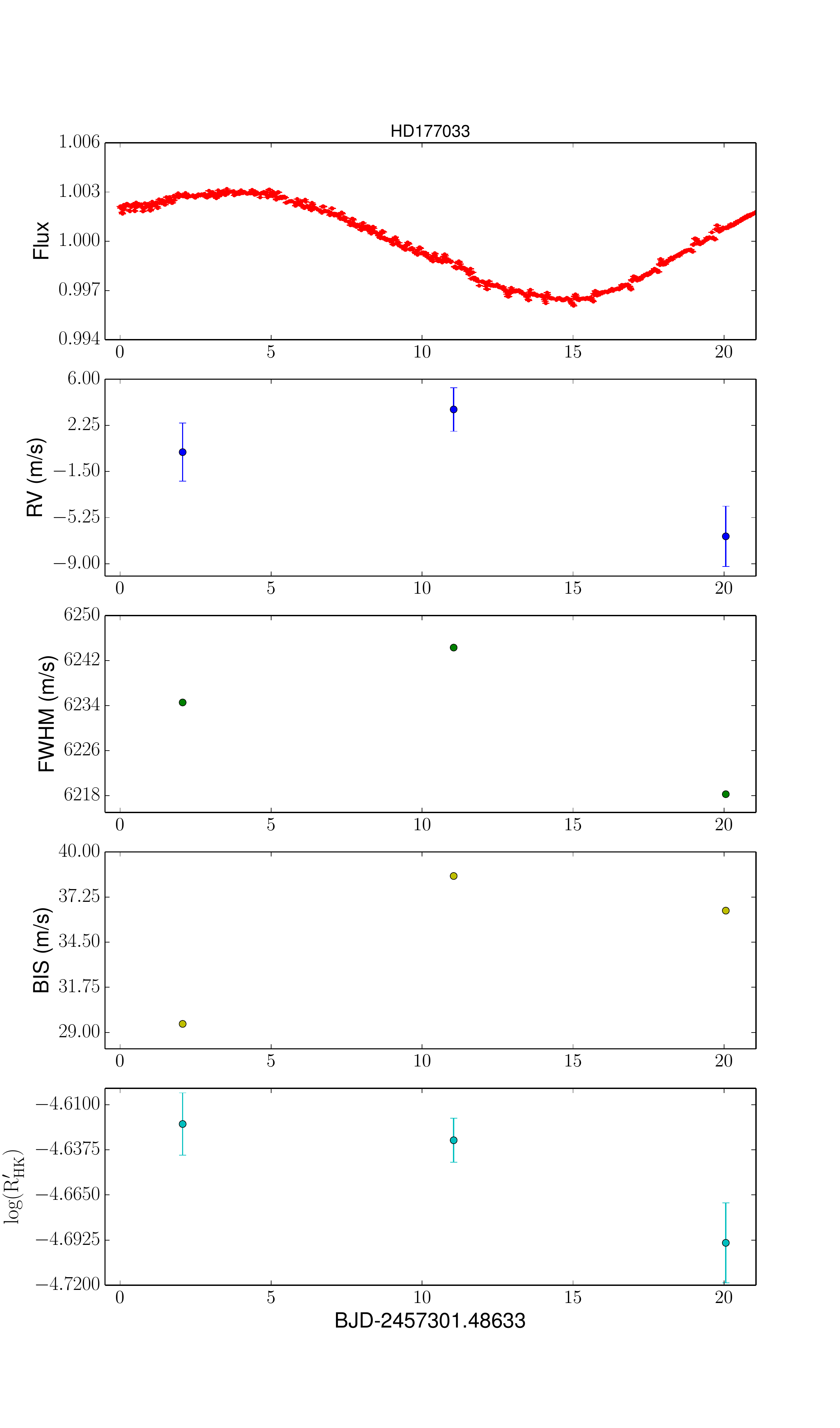}}\medskip\hspace{-0.2cm}\medskip 	\vspace{-2.cm}

\hspace{-1.3cm} \subfloat{\includegraphics[width=0.36\textwidth, height=90mm]{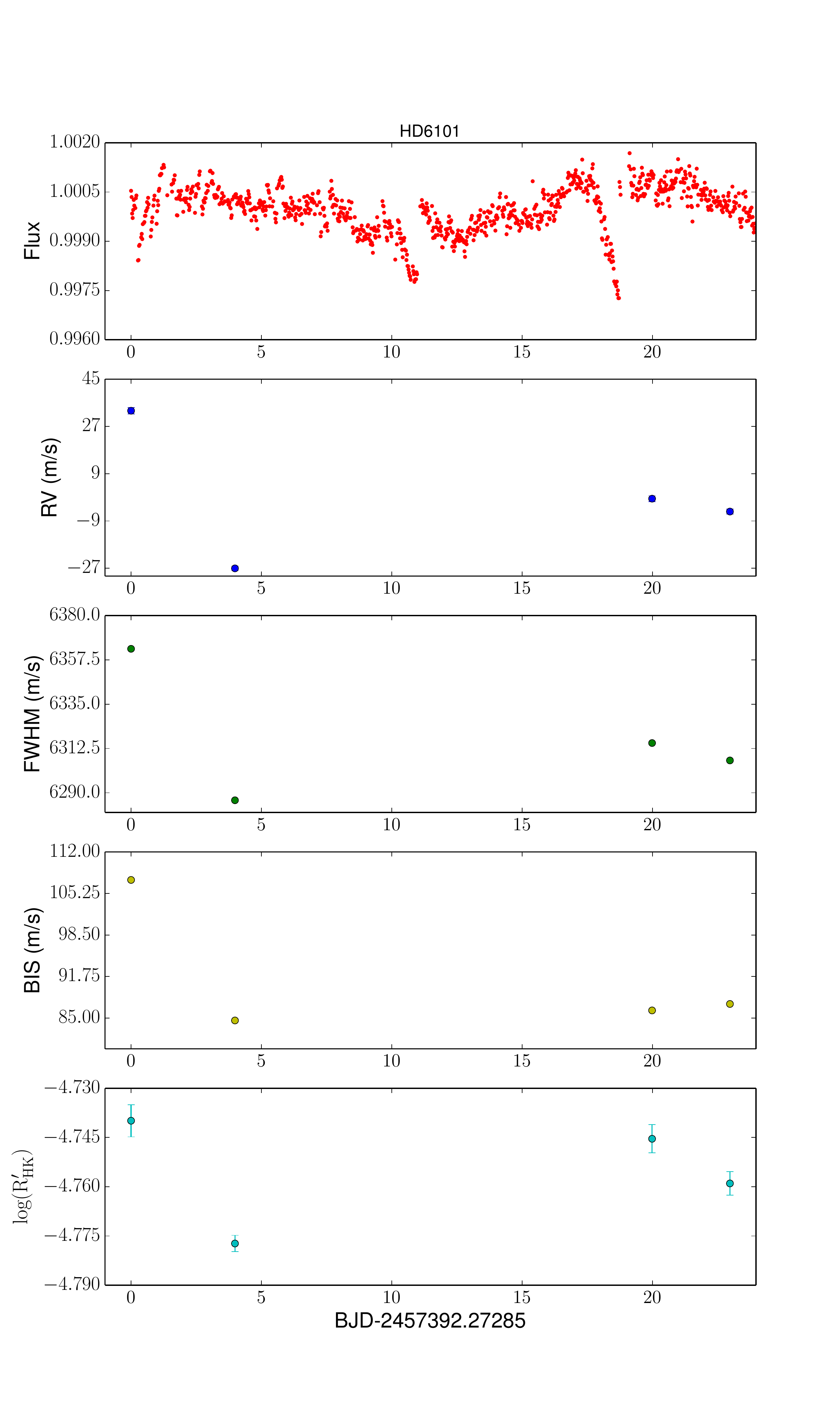}}\hspace{-0.2cm}\medskip
  \subfloat{\includegraphics[width=0.36\textwidth, height=90mm]{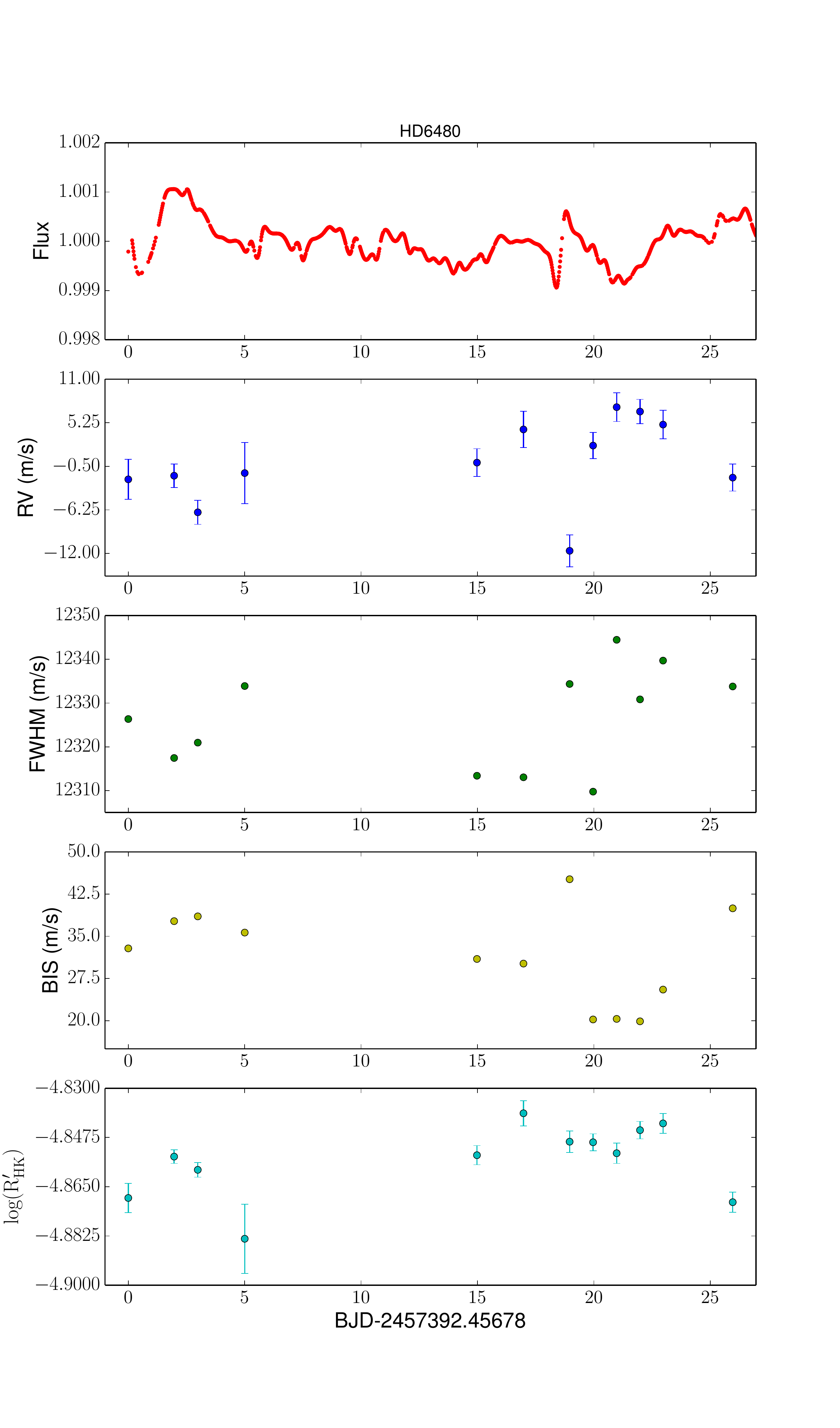}}\medskip\hspace{-0.2cm}\medskip
  \subfloat{\includegraphics[width=0.36\textwidth, height=90mm]{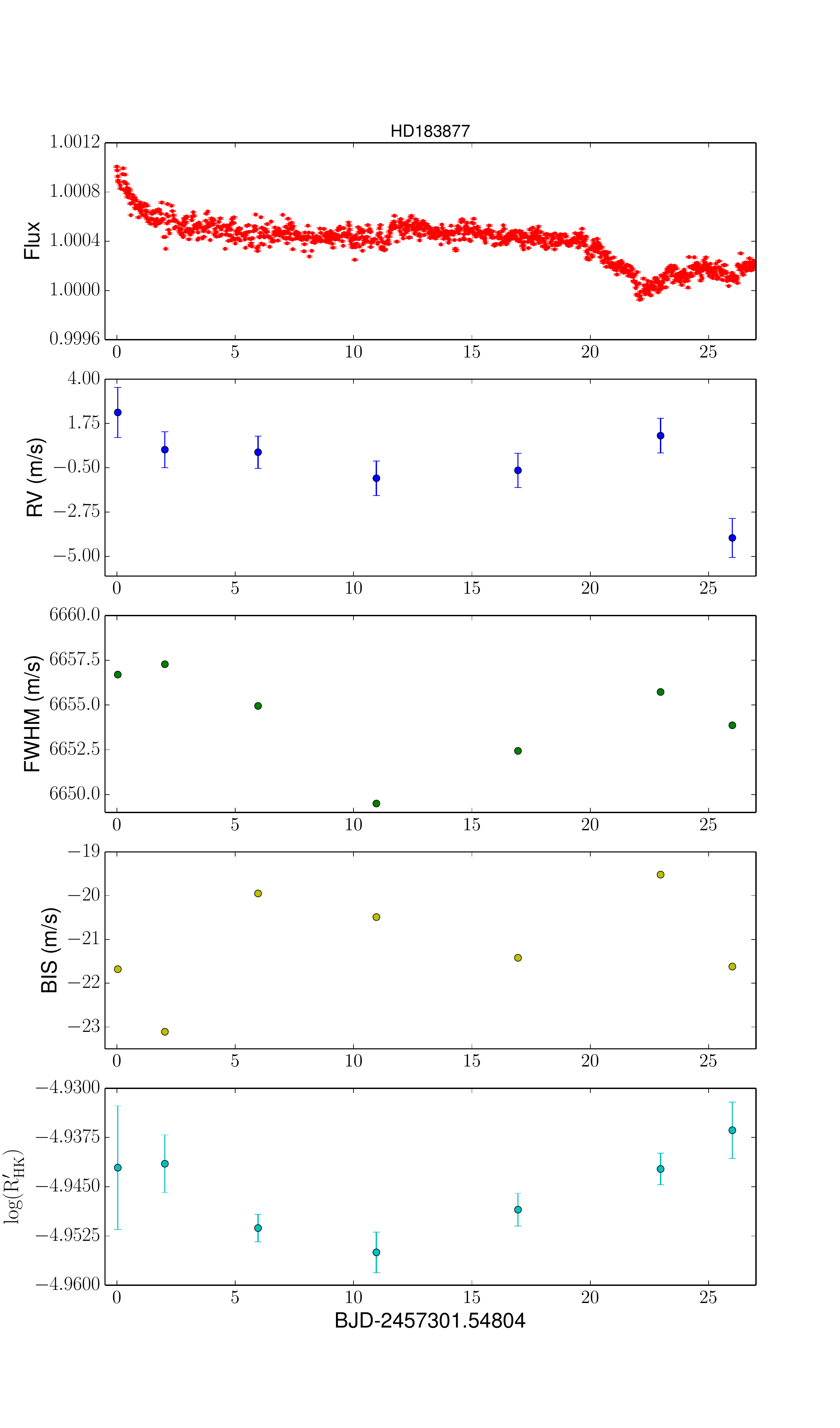}}\medskip\hspace{-0.2cm}\medskip \vspace{-2.cm}
  \hspace{-1.3cm} \subfloat{\includegraphics[width=0.36\textwidth, height=90mm]{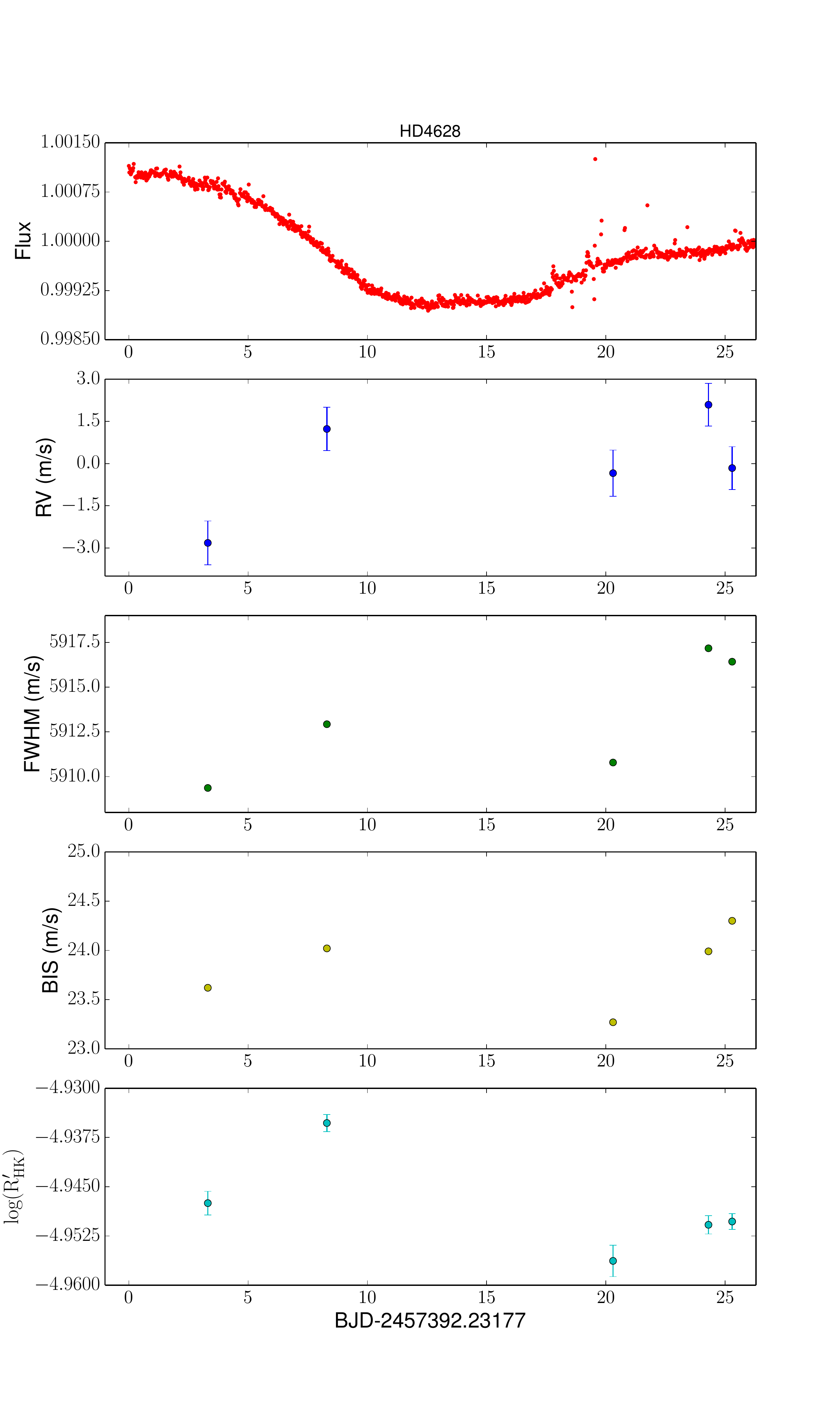}}\hspace{-0.2cm}\medskip
  \subfloat{\includegraphics[width=0.36\textwidth, height=90mm]{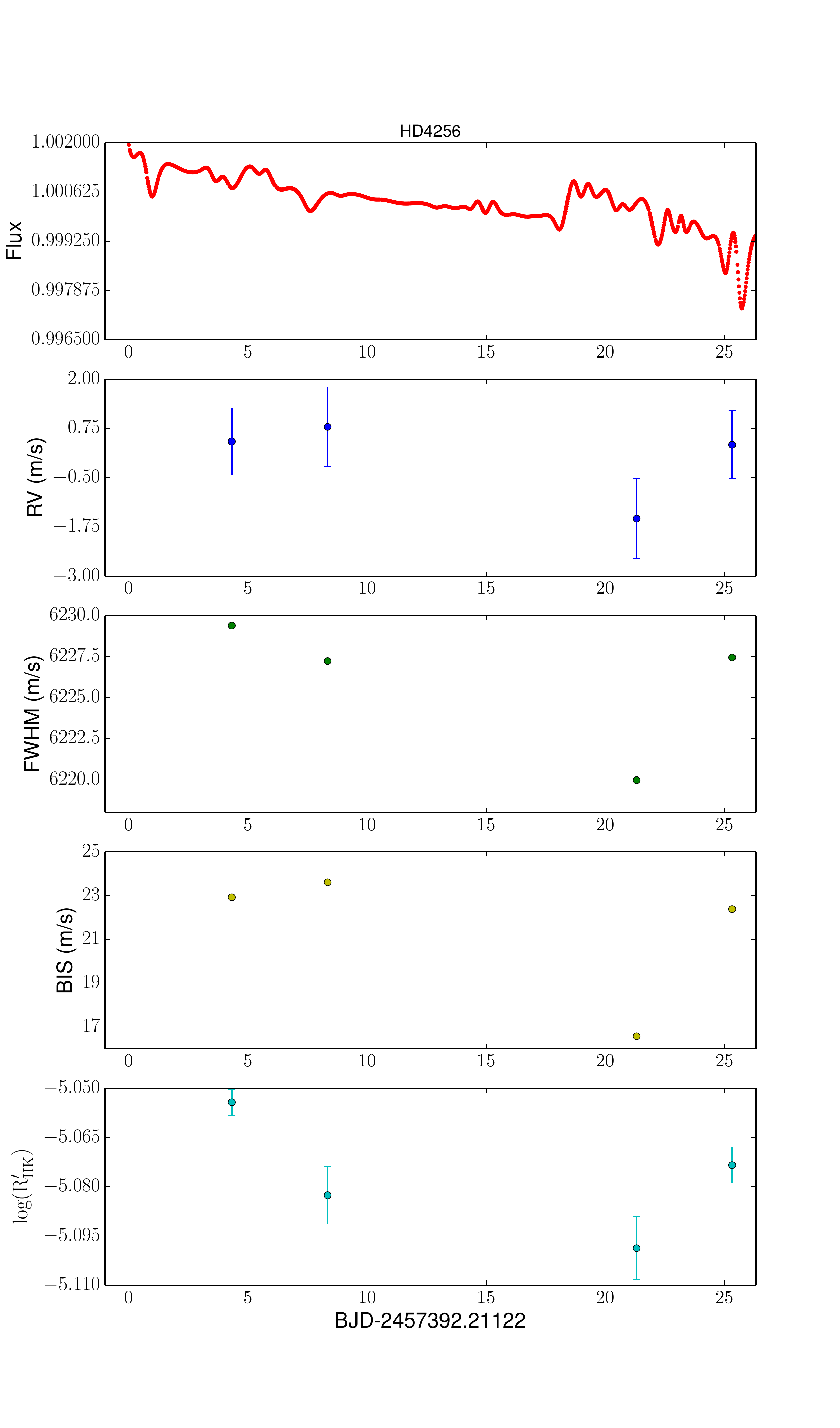}}\medskip\hspace{-0.2cm}\medskip \vspace{-1.cm}
  \caption{Simultaneous high-precision \textit{K2} photometric time series, RV, FWHM, BIS, and ($logR'_{HK}$) for each star in our sample. Sorted by increasing activity level based on $logR'_{HK}$. Note the different scales of photometric variability on each of the top panels. We would like to note that in some cases the errorbars are smaller then the symbol size.}

  \end{figure*}

We organize this paper as follows: in Section 2 we describe our observational datasets. In Section 3 we assess the correlation between the RV measurements, photometric variation, FWHM, BIS, $logR'_{HK}$, and 8-hr flicker. In Section 4, we examine and evaluate two main modeling approaches, namely \textit{FF'} \citep{Aigrain-12} and \textit{SOAP2.0} \citep{Dumusque-14}, in modeling activity induced RV jitter. In Section 5 we present the analysis of a star which we observed in the spectropolarimetric mode of HARPS in order to estimate its magnetic field of star and test for correlations with the RVs. Finally,
in Section 6 we summarize, and draw conclusions on, our results.

\section{Observations and data reduction}
\subsection{Target selection}
We created a list of potential targets by cross matching the stars that had an activity level characterized through the $logR'_{HK}$ index based on archival  HARPS spectra and the approved targets for Campaign 7 and 8 of \textit{K2}. From this, we tried to select stars with different activity levels, from very active  ($logR'_{HK} > -4.6$) to quiet ones ($logR'_{HK} < -4.8$). Our final sample contains nine stars. In Table 1 we list the main stellar parameters of our targets, where some of these parameters were obtained through the \textit{K2} Ecliptic Plane Input Catalog (EPIC) \footnote{https://archive.stsci.edu/k2/epic/search.php}.

\subsection{\textit{K2} data}

We downloaded the pixel data of all of our targets from the Mikulski Archive for Space Telescopes (MAST) \footnote{https://archive.stsci.edu/k2/}, and then utilized a modified version of the CoRoT imagette pipeline (known as POLAR) to extract the high-precision photometric time series. A full description of the POLAR pipeline was presented in \citep{Barros-16a}, and this pipeline has been used in several exoplanet discoveries from \textit{K2} \citep[e.g.][]{Barros-15, Armstrong-15, Lillo-Box-16, Santerne-16, Bayliss-17}. The stars analyzed in this study are very bright and were almost all saturated in the \textit{K2} data, therefore we adapted the apertures to include all the flux for the star \footnote{We extend the aperture at the bottom and the top of saturated columns to ensure that no photo-electrons smearing from the saturated pixels is lost throughout the campaign}. Furthermore, we check the co-detrending vectors to make sure that the variability in the light curves does not include instrumental effects.

\subsection{HARPS observations}
We were allocated three nights on the HARPS spectrograph, mounted on the ESO 3.6 m telescope at La Silla observatory  \citep{Mayor-03}, to carry out high-precision RV measurements of all of our targets (under ESO programme ID: 096.C-0708(A), PI: M. Oshagh). Thanks to the time-sharing scheme on HARPS with several other observing programs, we managed to spread our observations over 50 nights from the 5th of October 2015 to the 1st of November 2015, and from the 8th of January 2016 to the 3rd of February 2016.

The collected HARPS spectra were reduced using the HARPS Data Reduction Software (DRS -- \citealt{Pepe-02, Lovis-07}). In the DRS the spectra were cross-correlated with masks based on the target's stellar spectral type. As output the  DRS provides the RVs, FWHM of the cross-correlation function (CCF), BIS (as defined in
\citealt{Queloz-01}), $logR'_{HK}$ (as described in \citealt{Lovis-11}), and their associated uncertainties
following the methods described in \citet{Bouchy-01}.

We decided to observe one of our stars (HD179205) in the spectropolarimetric mode of HARPS (HARPS-Pol) instead of standard RV mode. Spectropolarimetric observations enable us to estimate the stellar magnetic field. Therefore, we exclude this star from our analysis in Sections 3 and 4; however, we dedicate a separate section (Section 5) for a detailed analysis of this star.

Figure 1 presents simultaneous high-precision \textit{K2} photometric time series, RV, FWHM, BIS, and ($logR'_{HK}$) for each of the eight stars in our sample. Note that the scale of photometric variability, which is shown in the top panels in all plots of Figure 1, covers a diverse range of values depending on each star's photometric variability. Therefore, one should be cautious when comparing the photometric variability of the stars to one-another.

\section{Probing the correlations between RVs and other observables}
\subsection{Flux, FWHM, BIS, and $logR'_{HK}$}
In this section we assess the presence of any meaningful correlation between all the observables versus the high-precision HARPS RV measurements, for each individual star. We inspect the presence of a correlation based on the Spearman's rank-order correlation coefficient ($\rho$)\footnote{Note that the Spearman's rank-order correlation assesses how well two variables can be described with a monotonic relationship and not purely linear.}. Moreover, we evaluate the significance of the measured $\rho$ using the straightforward Bayesian approach described in \citet{Figueira-16}. The posterior distribution of $\rho$ indicates what is the range of $\rho$ values that is compatible with the observations. In particular, it allows us to assess how probable a non-zero correlation value is.

Figure 2 presents the correlations between the RV measurement of each star and the corresponding photometric variation, FWHM, BIS, and $logR'_{HK}$. On the top right corner of each panel, the posterior distribution of $\rho$ and its most probable value are presented. The dashed vertical lines indicate the 68\% highest
posterior density credible intervals \citep{Figueira-16}. 

As one can notice for the case of very active stars, such as HD173427, there are strong (as reported by the high absolute value of the average $\rho$) and quite significant (as reported by its lower spread, narrower credible intervals, and incompatibility of the average value with zero) correlations between the measured RVs and all the simultaneously obtained observables, except the photometeric variability. As we move towards lower activity levels (as indicated by lower photometric variability, lower RV jitter rms, or smaller values of $logR'_{HK}$), the correlations become random. For instance, our result shows very strong correlation between the RV and BIS for the very quiet star (HD4256), and on the contrary shows no correlation between RV and BIS for the second most active star of our sample (HD181544). Therefore, from our results it is hard to conclude any trend or behaviour for the correlation between RV and other observables depending on the stellar activity level. In the quest for exoplanets, especially in RV surveys, quiet stars (as determined by low levels of photometric  variability or low level of activity determined based on their $logR'_{HK}$) are traditionally favored and targeted. However, our result suggests that the quiet stars might also be the challenging ones to disentangle their activity induced RV from the planetary RV signal, due to the lack of strong correlations between the RV jitter and any other observables. We list all the correlation coefficients and their 68\% highest
posterior density credible intervals in Table 2.


However, if we choose to determine the best activity indicator by exclusively inspecting carefully all the correlations presented in Table 2 and Figure 2, we can state that the FWHM appears as the best indicator. This is backed by the common idea that the correlation between FWHM and RVs (even-though not very strong or significant consistently) persists for stars of different activity levels. Our conclusion about the FWHM is in strong agreement with the findings of \citet{Queloz-09} and \citet{Pont-11}, who also suggested that the FWHM should be used
to reconstruct the RV jitter.

We would like to note that a phase shift is expected between the RV measurements and some of the other observables, due to geometrical or physical processes. For instance, \citet{Queloz-01} and \citet{Santos-03} were the first ones to pinpoint the presence of a phase shift between photometry and RV
signal. In addition, \citet{Queloz-09} estimated a phase shift of one quarter of the stellar rotation between FWHM and RV measurements of CoRoT-7. The presence of a phase shift might affect the correlation between two observables and we evaluate this in Appendix A. However, the low number of spectroscopic observations for each individual star in our sample prevents us from exploring any type of periodicity test on them as well as assessing the existence of phase-shift in them and also investigating its impact on our analysis. Moreover, it is worth mentioning that the inclination angle of the stellar rotation axis might also influence the presence and strength of the correlations between RVs and other observales although, since the stellar inclination of our targets are not known, exploring this possibility is beyond the scope of this paper.

\begin{figure*}[!h]
	\center

\includegraphics[width=\textwidth, height=228mm]{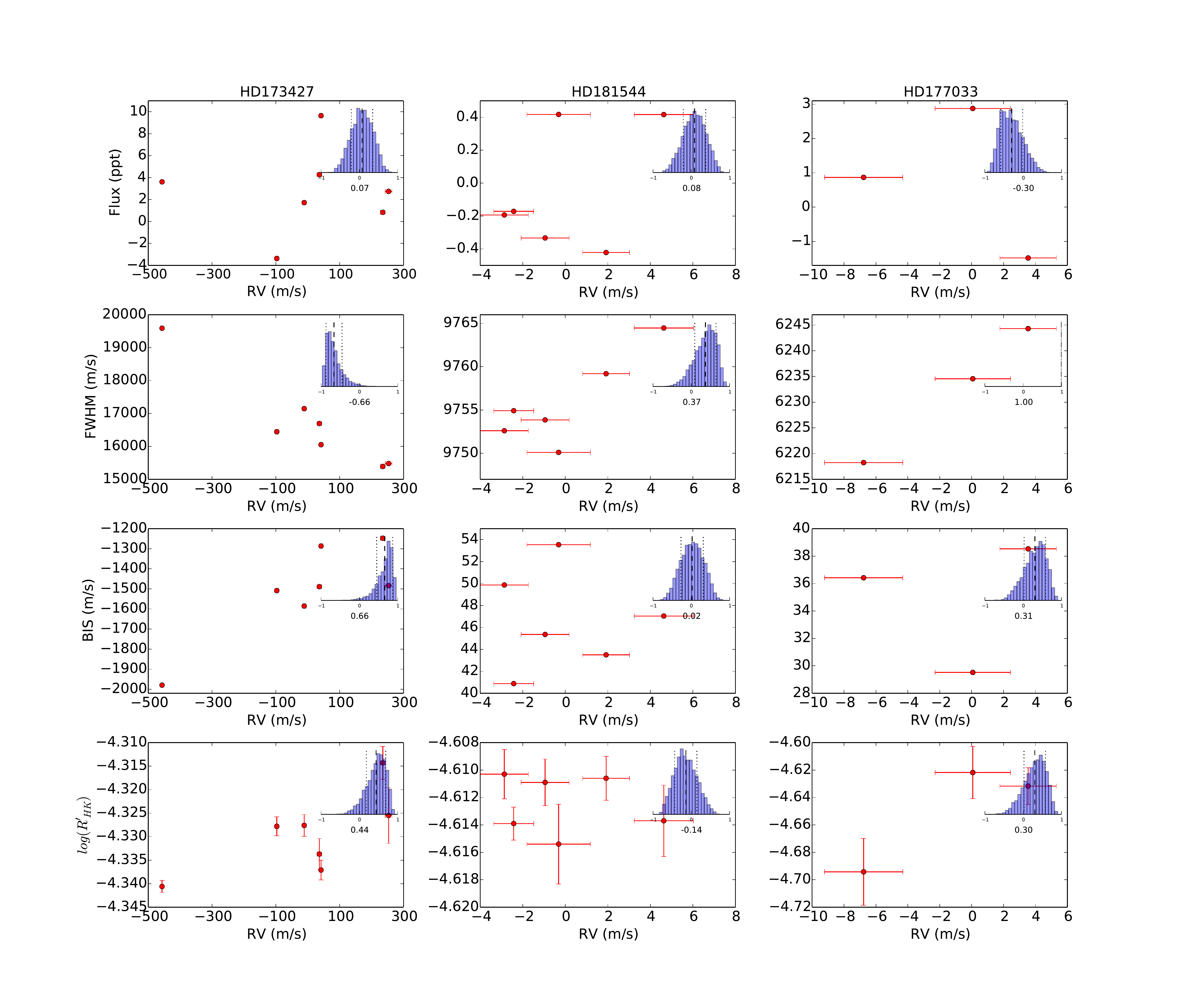}
\caption{Correlations between the RV measurements and the corresponding
photometric variations, FWHM, BIS, and $logR'_{HK}$. Sorted by increasing activity level based on $logR'_{HK}$. On the top right of each panel the calculated value of $\rho$ and also its posterior distribution is presented. The dashed vertical lines indicate the 68\% highest
posterior density credible intervals \citep{Figueira-16}. }
\end{figure*}

\begin{figure*}[!h]
	\center

\ContinuedFloat
\includegraphics[width=\textwidth, height=228mm]{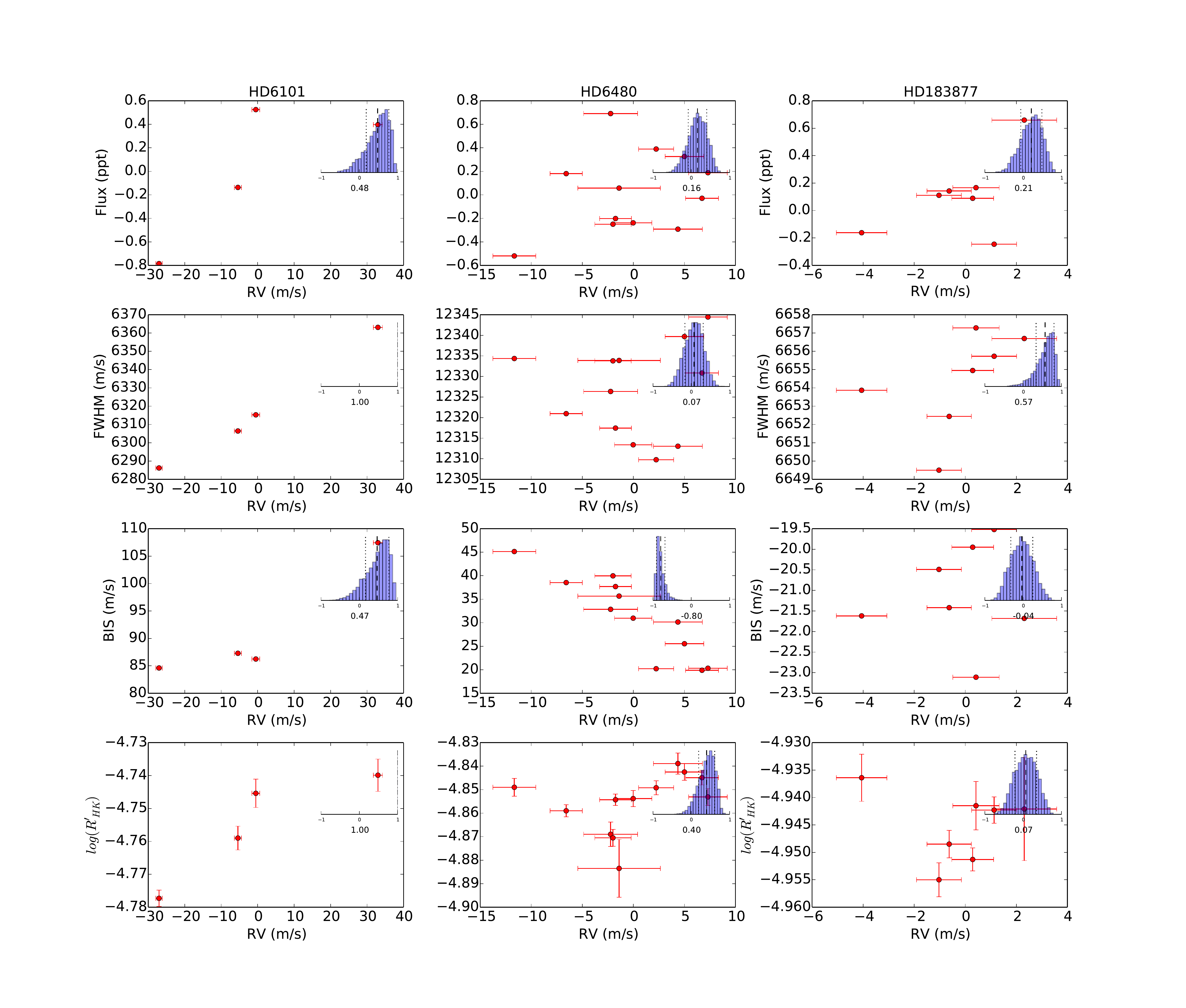}
  \caption{Continued.}
\vspace{2mm}

\end{figure*}
\begin{figure*}[!h]
	\center

\ContinuedFloat
\includegraphics[width=\textwidth, height=228mm]{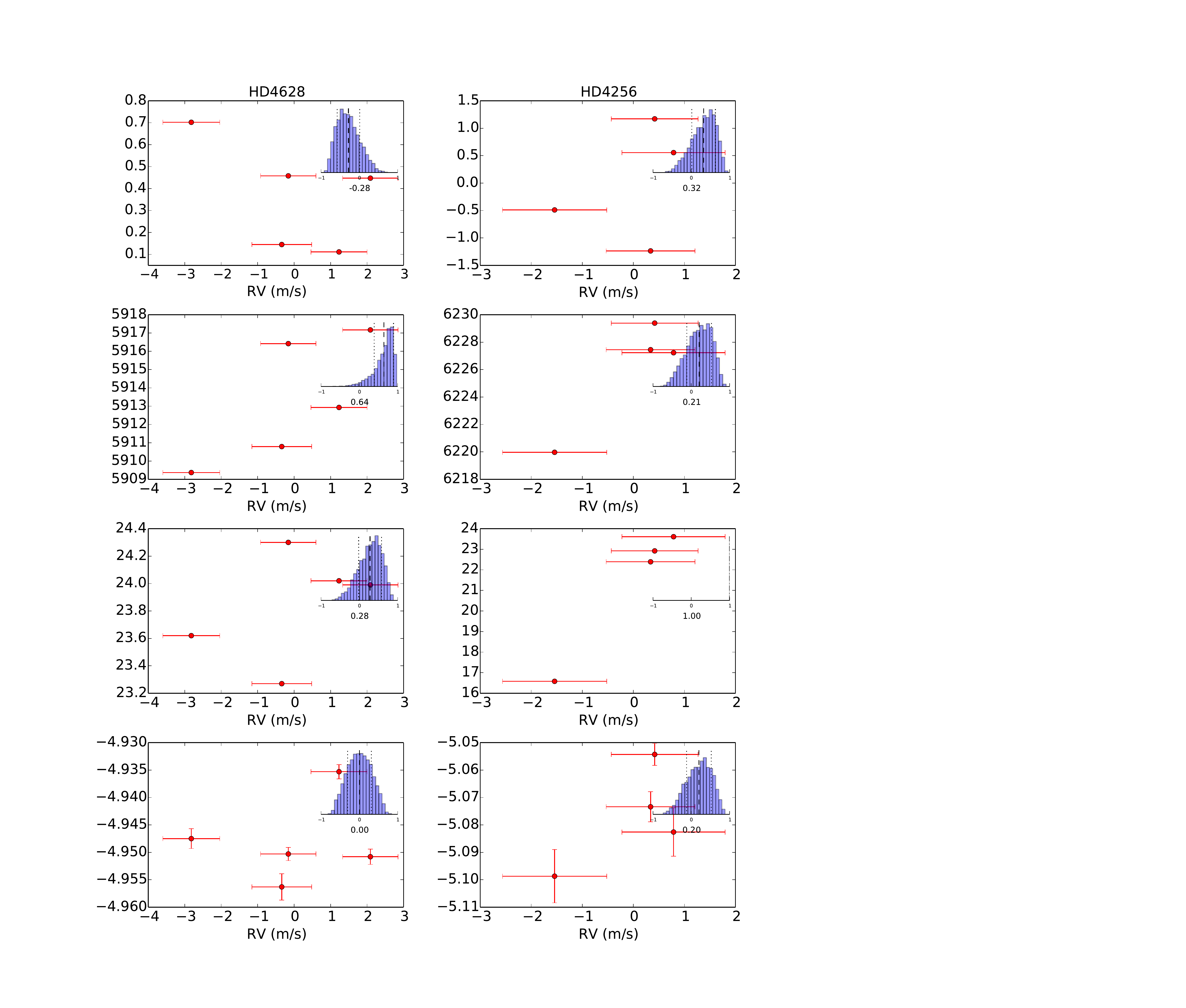}
    \caption{Continued.}
\vspace{2mm}

\label{fig:1}
\end{figure*}

\begin{table*}
\caption{Calculated Spearman's rank-order correlation coefficients and their 68\% highest
posterior density credible intervals. Sorted by increasing activity level based on $logR'_{HK}$.}              
\label{table:1}      
\centering                                      
\begin{tabular}{c c c c c c}          
\hline\hline  
\\                      
Name & RV vs. FLUX & RV vs. FWHM & RV vs. BIS & RV vs. $logR'_{HK}$ & $logR'_{HK}$\\    
\hline   
\\                                
HD173427 & 	\textbf{0.07} $\pm$ 0.28 & \textbf{-0.66} $\pm$ 0.20 & \textbf{0.66} $\pm$ 0.20 & \textbf{0.44} $\pm$ 0.31 &-4.33\\
HD181544 & 	\textbf{0.08} $\pm$ 0.30 & \textbf{0.37} $\pm$ 0.27 & \textbf{0.02} $\pm$ 0.30 & \textbf{-0.14} $\pm$ 0.27&-4.61\\
HD177033 & 	\textbf{-0.30} $\pm$ 0.29 & \textbf{1.00} $\pm$ 0.0001 & \textbf{0.31} $\pm$ 0.28 & \textbf{0.30} $\pm$ 0.28&-4.66\\
HD6101 & 	\textbf{0.48} $\pm$ 0.30 & \textbf{1.00} $\pm$ 0.0001 & \textbf{0.47} $\pm$ 0.29 & \textbf{1.00} $\pm$ 0.0001&-4.76\\
HD6480 & 	\textbf{0.16} $\pm$ 0.23 & \textbf{0.07} $\pm$ 0.21 & \textbf{-0.80} $\pm$ 0.10 & \textbf{0.40} $\pm$ 0.21&-4.86\\
HD183877 & 	\textbf{0.21} $\pm$ 0.28 & \textbf{0.57} $\pm$ 0.23 & \textbf{-0.04} $\pm$ 0.28 & \textbf{0.07} $\pm$ 0.25&-4.94\\
HD4628 & 	\textbf{-0.28} $\pm$ 0.29 & \textbf{0.64} $\pm$ 0.24 & \textbf{0.28} $\pm$ 0.30 & \textbf{0.00} $\pm$ 0.29&-4.94\\
HD4256 & 	\textbf{0.32} $\pm$ 0.31 & \textbf{0.21} $\pm$ 0.30 & \textbf{1.00} $\pm$ 0.0001 & \textbf{0.20} $\pm$ 0.25&-5.08\\

\hline                                             
\end{tabular}
\end{table*}

\subsection{F8}
The term ``8-hr flicker" or simply ``F8", which is defined as the RMS of photometric observations on time scales shorter
than 8 hours, was introduced by \citet{Bastien-13}. \citet{Bastien-13} found that the F8 values exhibit a strong correlation
with the surface gravity of the stars determined asteroseismically and, therefore, proposed the idea of using the F8 value to estimate the stellar surface gravity. Moroever, \cite{Bastien-14} also identified a tentative correlation between the F8-based stellar surface gravity and the RV jitter. Furthermore, \citet{Cegla-14} explored the possible correlation between the F8 and RV jitter and also between the F8 diagnostic and photometric variability, and also provided evidence that a temperature sensitive correlation exists for quiet stars. In this section we aim to assess possible correlations between the measured F8s and the RV jitter amplitude and also between F8 and the photometric variation amplitude.

We calculate the F8 value for all our targets according to the description provided by \citet{Bastien-13}. As we mentioned in the introduction \citet{Bastien-14} performed a similar study using high-precision RV measurements, obtained from Keck and Lick observatories, and high-precision photometric observations of 12 \textit{Kepler} stars. Although their RV measurements were not taken simultaneously with \textit{Kepler}'s photometric observations, their stars do cover quite a wide range of stellar activity (though most are in the low activity regime)\footnote{Note that as mentioned in \citet{Bastien-14} the RV jitter presented in \citet{Bastien-14} with values less
than $4 {\rm m\,s}^{-1}$ may be dominated by instrumental systematics
and shot noise. Therefore, such values are considered as upper
limits.}. Therefore, here we decided to combine their stars with our sample and study them together \footnote{We would like to mention that the F8 values for the \citet{Bastien-14} stars were obtained via private communication because the values published in \citet{Bastien-16} catalog for these stars were erroneous (Bastien private communication). An erratum of \citet{Bastien-16} catalog is under preparation to be submitted.}.

Figure 3 presents the correlation between the measured F8 and their corresponding RV jitter and photometric variation amplitudes. We examined the correlation between the variables, similar to the section 3 and Figure 2, by evaluating the posterior distribution of Spearman's correlation coefficient $\rho$ and its most probable value are presented. As Figure 3 shows, the F8 values of stars in our sample exhibit strong and significant correlations with the RV jitter and photometric variation amplitude (shown as the filled circles, green histograms and green values in Figure 3). When we combine our targets with \citet{Bastien-14} and evaluate the total sample the correlation between F8 and RV jitter becomes stronger and more significance, and on the other hand the correlation between F8 and photometric variation becomes much weaker and also less significant. Therefore, our results indicate that F8 might be a powerful proxy for activity induced RV over a wide range of stellar activity. However, we would like to note that \citet{Bastien-14} stars were mostly from the low activity regime (based on their low photometric variability and also RV jitter amplitude) and a lack or presence of correlation in the low active regime may have diluted or strengthened the total correlation. Moreover, the \citet{Bastien-14} RV measurements were not taken simultaneously with \textit{Kepler}'s photometric observations, this delay would lead to an artificial correlation in the \citet{Bastien-13} sample, and thus could bias the total correlation. 

We also attempted to evaluate the dependence of correlations on the stellar temperatures, as presented by the color-coded points in Figure 3. However, we could not identify any meaningful trend between the correlations and stellar temperatures.

\begin{figure}[t!]

\includegraphics[width=0.5\textwidth, height=110mm]{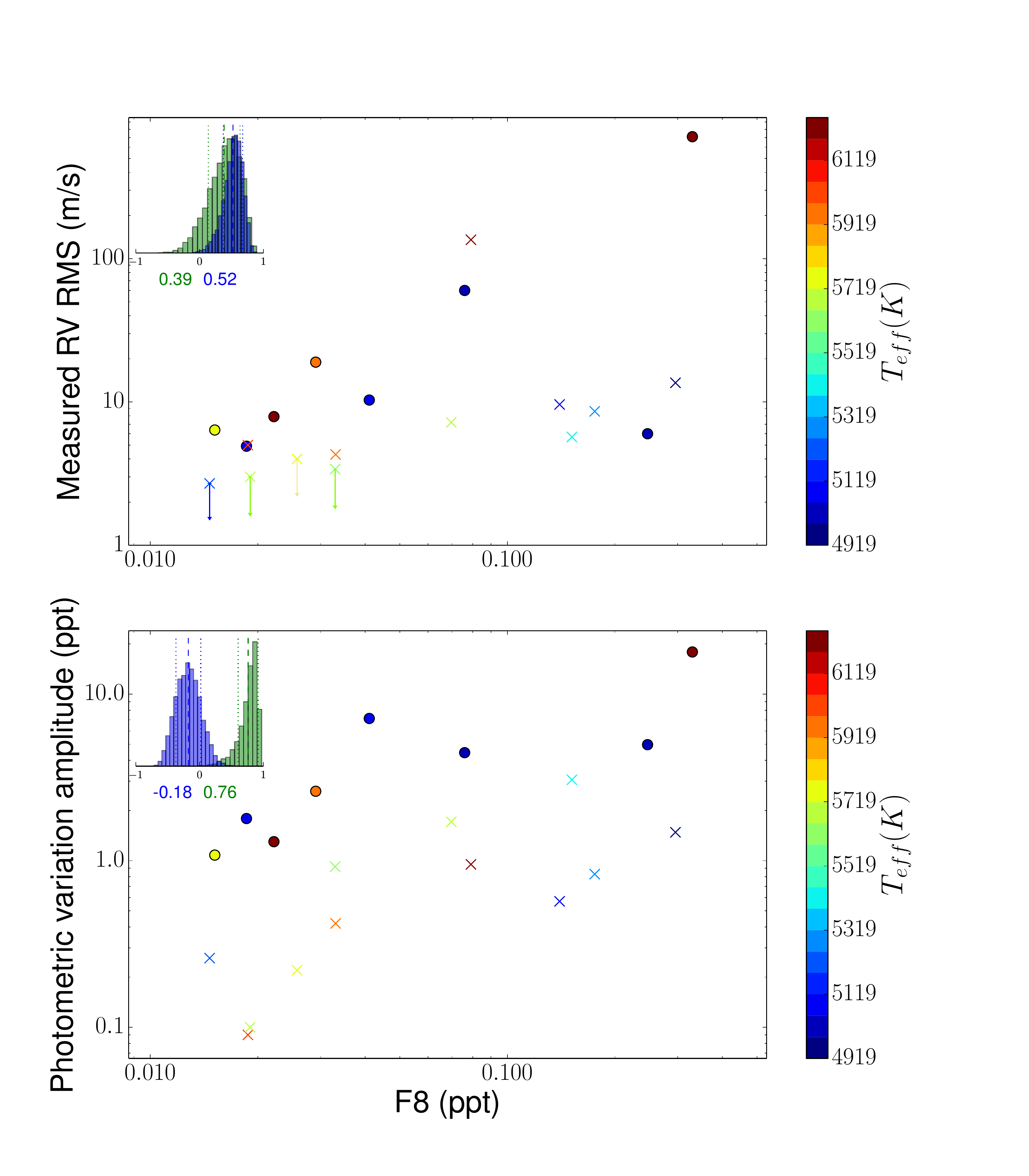}

 \caption{Top: Correlation between measured F8 values and RV jitter amplitude of our targets (represented by filled circles) and also the stars from \citet{Bastien-14} (displayed by crosses). On the top left the calculated value of $\rho$ and also its posterior distribution for our targets (green histogram and green value) and also for the full sample are presented (blue histogram and blue value). The RV jitter values reported in \citet{Bastien-14} with values less
than $4 {\rm m\,s}^{-1}$ are considered as the upper
limits, and are indicated by descending arrows in this plot. The points are color-coded according to their $T_{eff}$. Bottom: same as the top but for the F8 and photometric variation amplitude. }
  \label{sample-figure}
\end{figure}

\section{Modeling}
In this section we aim at assessing how accurately one can predict the RV jitter amplitude using the simultaneous high-precision photometric time series and by applying two different modeling techniques, namely \textit{FF'} and \textit{SOAP2.0}, across different levels of stellar activity.

\subsection{The \textit{FF'} method}
The \textit{FF'} method, which was introduced by \cite{Aigrain-12}, provides a simple technique to predict the RV jitter from simultaneous photometric time series and its first derivative. This method considers two main components: a rotational one, and another arising from convective blue-shift variations. The rotational term estimates the RV due to the contrast of active regions
on the rotating star, and can be shown to be given by, as in \citet{Aigrain-12},

\begin{equation}
\Delta RV_{rot}(t)= -\frac{\dot \psi(t)}{\psi_{0}} \left[1-\frac{\psi(t)}{\psi_{0}}\right] \frac{R_{\star}}{f},
\end{equation}

\noindent where $\psi(t)$ is the flux, and $\dot \psi(t)$ is its first derivative. $R_{\star}$ is the stellar radius in solar radii, $\psi_{0}$ is the maximum of $\psi(t)$ when there are no spots on the visible part of the star, and $f$ is the reduction in the flux which is equal to the fractional area of a spot to the stellar disk.




The convective blue-shift term takes into account the impact of suppression of convective blue-shift in the active region's area, which can be formulated as, as in \citet{Aigrain-12},

\begin{equation}
\Delta RV_{CB}(t)= \left[1-\frac{\psi(t)}{\psi_{0}}\right]^{2}\frac{\delta V_{c} \kappa}{f}
\end{equation}

\noindent where $\delta V_{c}$ is the difference of the convective blueshift velocity between inactive and active regions. $\kappa$ is the ratio of the unspotted area to the active region area (typically $\gg 1$).

Therefore, the total activity-induced RV prediction of the \textit{FF'} method can be estimated as,

\begin{equation}
\Delta RV_{activity}(t)= \Delta RV_{rot}(t)+  \Delta RV_{CB}(t).
\end{equation}

We compared the amplitude of the \textit{FF'} predicted RV jitter to the amplitude of the observed RV jitter for each individual star in our targets list. To do this we first smoothed the photometric observations (using an iterative nonlinear filter \citet{Aigrain-04}) and calculated its first derivative for each star. We adopted the stellar radius reported in Table 1 for Eq 1. Since the exact values of $f$, $\delta V_{c}$, and $ \kappa$ are not known they are treated as free parameters. We obtain their value by fitting the maximum and minimum of the \textit{FF'} predicted RVs to the maximum and minimum of the observed RVs, respectively, using an Levenberg-Marquardt algorithm. We would like to mention that these parameters had some boundaries, due to their definition and physical constraint. For instance, $f$ has to be in the range of [0,1], therefore, we set those boundaries to prevent the fitting procedure from attempting to obtain non-physical values.

\subsection{The \textit{SOAP} tool}
\textit{SOAP} is a publicly available tool that simulates the effects of dark spots and bright plages on the surface of a rotating star by taking into
account the flux contrast effect \citep{Boisse-12}. \textit{SOAP 2.0} is an upgraded version of \textit{SOAP} \footnote{The tool is publicly accessible through www.astro.up.pt/resources/soap2/}. \textit{SOAP 2.0} can generate
the photometric and RV variation signals
induced by active regions by taking into
account not only the flux contrast effect in those regions,
but also the RV shift due to inhibition of the convective
blueshift inside those regions \citep{Dumusque-14}. \citet{Dumusque-14} demonstrated that the RV jitter of a spot-dominated star is dominated by the temperature contrast of spots combined with the stellar rotation; however, for the case of a plage-dominated star the inhibition of convective
blueshift takes over. Therefore, we could presume that the \textit{SOAP2.0} tool can be considered as the numerical equivalent of the FF' approach, attempting to simulate the two main effects (flux contrast and suppression of convective blueshift).

\textit{SOAP 2.0} allows us to predict the amplitude of RV jitter from the photometric time series. To do so, first we adjust all the required stellar parameters in \textit{SOAP2.0} according to their value from Table 1. We assume a single spot\footnote{We note that since we are interested in modeling the total amplitude of photometric variability and RV jitter, the inclusion of more spots/plages with total filling factor as a single one (to generate the observed amplitude) just broaden the shape of photometric and RV signal and does not affect the amplitude of those signals, and thus, does not provide extra information.} on the stellar surface with a fixed temperature contrast of -663 K, corresponding to the average temperature contrast between a sunspot and quiet region on the Sun as determined by \citet{Meunier-10}. Then we vary the spot's size until the photometric output of \textit{SOAP2.0} matches the amplitude variation of the observed light curve. At that point we compare the predicted RV jitter from \textit{SOAP2.0} to the observed one. We repeated the same procedure assuming a single plage with a temperature contrast of +250 K on the stellar surface instead of a spot. \citet{Meunier-10} showed that the temperature contrast of a plage located at the solar disc centre was +250 K. We would like to note that due to the smaller temperature contrast of plages compared to those of spots, the required plage filling factor needed to generate same photometric variability as spots is usually much larger, often by a factor of four, than the spot-filling factor. This photometric constraint has clear implications: due to larger areas in which the inhibition of convective blueshift occurs, a plage results in a much larger RV jitter amplitudecompared to a single spot.

\subsection{Comparison of predicted and observed RV jitter}
In this section we assess the ability of the \textit{ FF'} and \textit{SOAP2.0} methods in predicting the amplitude of RV jitter from simultaneous photometry time series.  Figure 4 presents the amplitude of predicted over
observed RV jitter as a function of photometric variability (shown in the markers with circle around).

As we mentioned, earlier \citet{Bastien-14} performed a similar study using high-precision RV measurements, obtained from Keck and Lick observatories, and high-precision photometric observations of 12 stars observed by \textit{Kepler}. Even though their RV measurements were not taken simultaneously with \textit{Kepler}'s photometric observations, their choice of stars covered a wider activity value range towards the lower activity regime. Therefore, we decided to repeat the procedures presented in Sections 4.1 and 4.2 on their observations and overplot the result for their targets in Figure 4 (shown by the markers without surrounding circles).

As Figure 4 presents, it is clear that the \textit{FF'} method systematically underpredicts the RV jitter amplitude for the whole range of photometric variabilities considered, and particularly for the low activity stars (as measured by their photometric variability). This finding is in strong agreement with the \citet{Bastien-14} result. The accuracy of the \textit{SOAP2.0} models strongly depend on the level of photometric variability. For instance, \textit{SOAP2.0} predicions arising from the single spot model improves significantly when the photometric
variability reaches above 1 ppt limit. On the contrary for quiet stars, when the photometric variability is smaller than 1 ppt, the \textit{SOAP2.0} single plage model provides more accurate predictions while overestimating the RV jitter for photometrically active stars. Thus, our results can be interpreted as the confirmation of the presence of two distinctly different activity regimes, spot-dominated and plage-dominated ones, and are in strong agreement with studies such as \citet{Lockwood-97},\citet{Radick-98}, and \citet{Shapiro-16}, which suggested a transition between spot-dominated activity for stars at higher activity levels to plage-dominated activity at lower activity levels.

\begin{figure}[t!]

\includegraphics[width=0.5\textwidth, height=70mm]{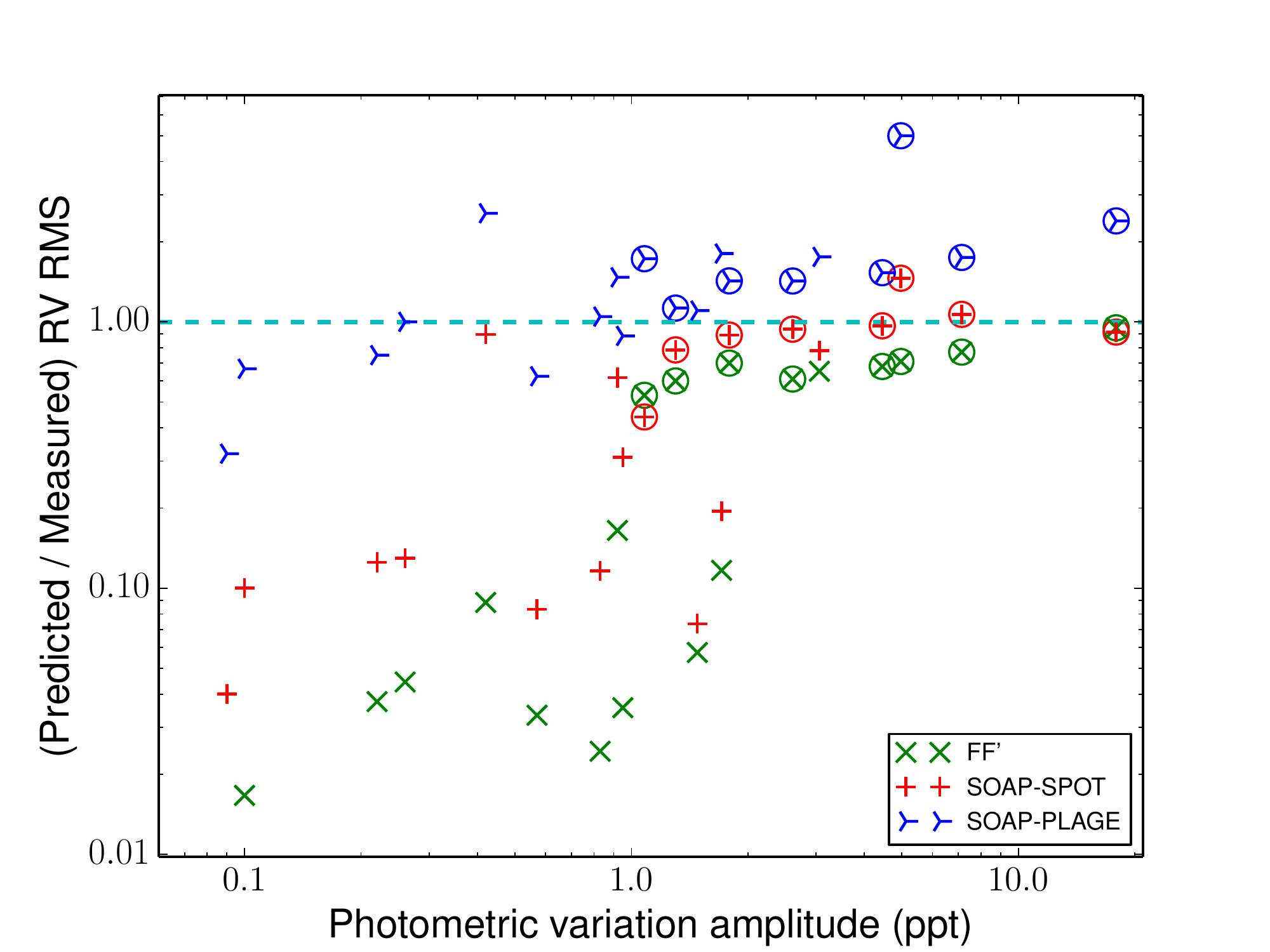}

 \caption{The predicted versus
observed RV rms as a function of photometric variability. The green "x"'s denote the \textit{FF'} method predictions, red pluses show \textit{SOAP2.0} single spot model, and blue triangles (right) displays the \textit{SOAP2.0} single plage model. The stars from our sample are presented by the markers enclosed by circles and the results from analyzing stars from \citet{Bastien-14} are presented by the markers lacking surrounding circles.}
  \label{sample-figure}
\end{figure}

\section{Spectropolarimetric observation of HD179205}

As mentioned earlier, one of our targets (HD179205) was observed in the spectropolarimetric mode of HARPS (HARPS-Pol), instead of standard RV measurements, in order to enable us to measure the line-of-sight-projected magnetic field averaged over the visible hemisphere of the star (also called longitudinal field or $B_{l}$). Although, $B_{l}$ is no more than a first-order magnetic proxy unable to capture the full complexity of the field topology, it can still be used to get a first hint on the large-scale field at the surface of the star when Stokes V signatures are reliably detected at the surface of the star.  

We first tried to estimate accurately the stellar rotation period of HD179205. To do that we utilized the entire \textit{K2} photometric time series of HD179205 and then applied the Generalised Lomb-Scargle periodogram \citep{Zechmeister-09}. Figure 5 presents the whole \textit{K2} light curve and also the resulting periodogram, which exhibits a significant peak at 5.047 days. Thus, we estimated $P_{rot}= 5.047 \pm 0.15$ days (the error was estimated from the full width at half maximum of the peak in the periodogram) which indicates that HD179205 is a moderately rapidly rotating star. In order to ensure that our estimated value for the stellar rotation is accurate and it is not its first harmonic $P_{rot}/2$, we used the relation provided by \cite{Noyes-84} to estimate the stellar rotation based on the stellar B-V and log $logR'_{HK}$. Assuming B-V = 0.55 and log $logR'_{HK}$ = -4.55, we estimated $P_{rot}= \sim 5.9$ days, which is in broad agreement with our estimate.

In the spectropolarimetric mode, each spectropolarimetric observation consists of a sequence of four subexposures recorded with the polarimeter quarter-wave plate set to different pre-selected azimuths.  By applying Least-Squares Deconvolution (LSD) \citep{Donati-97} to the observed spectra, we obtained Stokes I and V LSD profiles, allowing us to estimate the $B_{l}$ at each observing epoch. Despite reaching error bars as low as a few G on $B_{l}$, the field of HD179205 is not reliably detected in any of our observing epochs. This is not surprising given the spectral type and rotation period of the star, for which typical $B_{l}$ values of no more than a few G are to be expected, thus requiring sub G error bars on $B_{l}$ for a definite detection.  

Figure 6 shows the simultaneous \textit{K2} high-precision photometric and longitudinal magnetic field estimation and RV variations. Our results demonstrate how difficult it is to detect and estimate $B_{l}$ with enough precision for RV filtering studies.

Future studies with nIR spectropolarimeters like SPIRou at CFHT should be able not only to reach improved precisions on $B_{l}$ (thanks to the larger Zeeman splitting at infrared wavelengths) but also to detect the Zeeman broadening of spectral lines giving access to the unsigned magnetic flux at the surface of the star, thought to be one of the best proxies for the RV activity jitter \citep{Haywood-16}.

\begin{figure}[t!]

\includegraphics[width=0.45\textwidth, height=75mm]{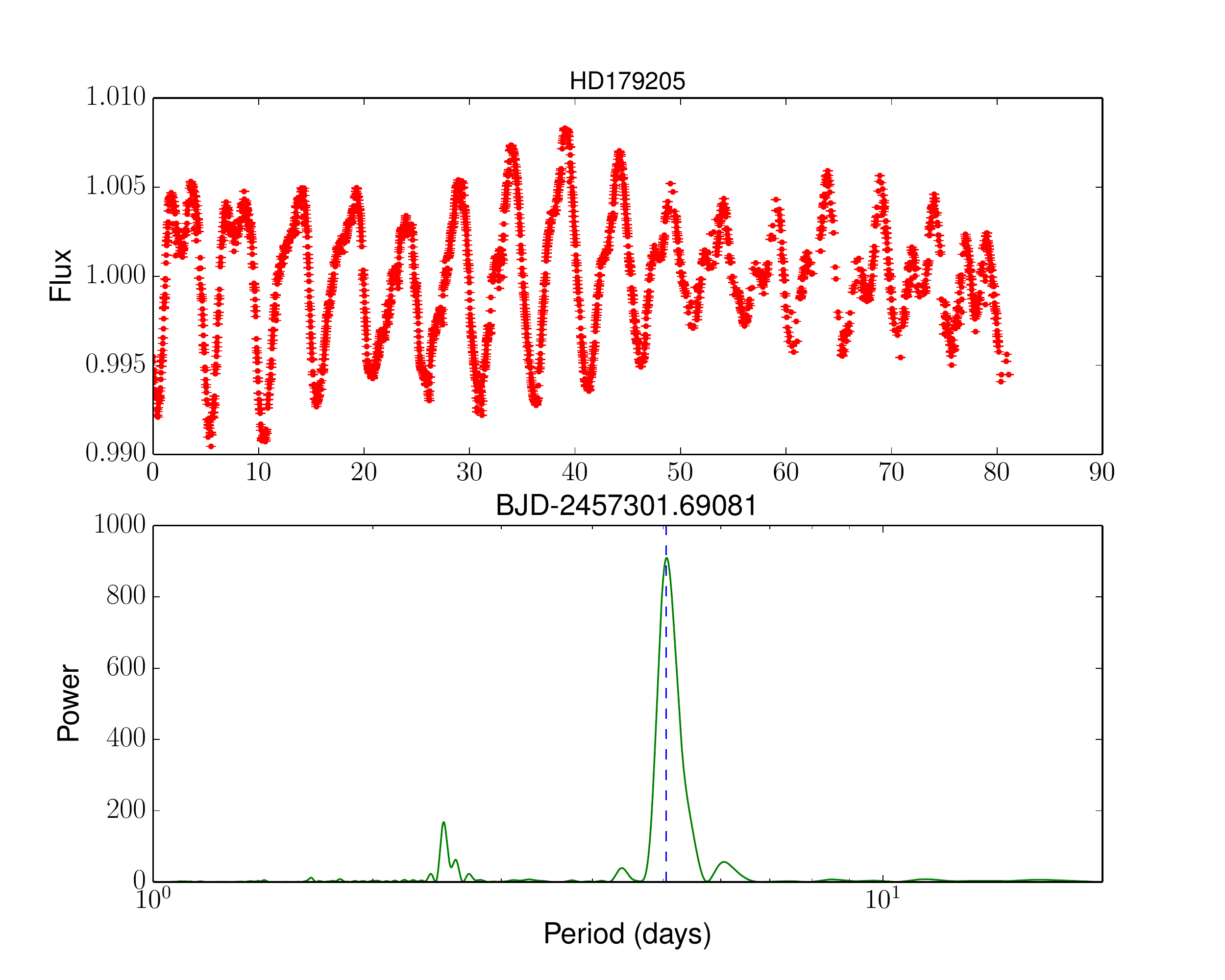}

 \caption{Top: The entire \textit{K2} photometric time series of HD179205 spanning almost 80 days. Bottom: the GLS periodogram of the photometric time series. The dashed line indicates the exact value of the peak of the periodogram corresponding to the stellar rotation period.}
  \label{sample-figure}
\end{figure}

\begin{figure}[t!]
\includegraphics[width=0.45\textwidth, height=105mm]{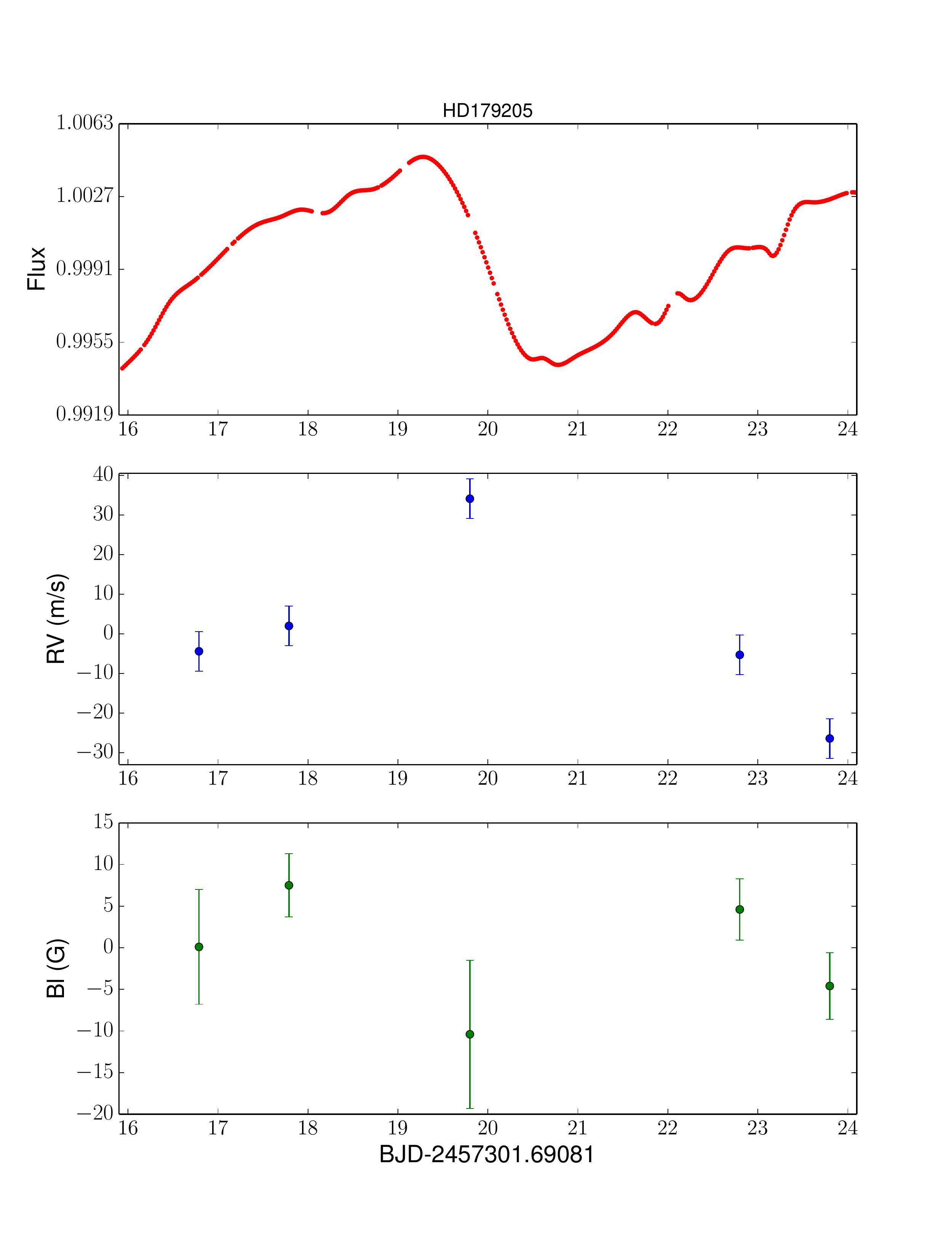}

 \caption{Simultaneous high-precision \textit{K2} photometric time series, RV, and $B_{l}$ of HD179205.}
  \label{sample-figure}
\end{figure}

\section{Conclusion}
In this paper we present simultaneous high-precision photometric observations (taken with the \textit{K2} space telescope) and RV measurements (obtained with the HARPS spectrograph) for a sample of nine stars. We first assessed the presence of any meaningful relationships between the measured RV jitter and the simultaneous photometric variations, as well as with other activity indicators (such as BIS, FWHM, and $logR'_{HK}$), by evaluating the strength and significance of the correlation between each of the observables.

We found that strong and significant correlations exist between almost all the observables and the measured RVs for the very active star in our sample and, therefore, they can (in principle) all be used as a proxy for activity. However, as we move towards lower stellar activity levels the correlations become random, and we could not reach any conclusion regarding the tendency of correlations depending on the stellar activity level. Traditionally, quiet stars are favored targets for exoplanet searches, however, our result suggests that they are challenging ones in which to disentangle their activity induced RVs from the planetary RV signal. We further calculated the F8 values of all our targets, and illustrated that F8 exhibits a distinct and strong correlation with the RV jitter, and not strong correlation with the photometric variability amplitude. Therefore, our results indicated that F8 could be used as a good proxy
for estimating the amplitude of RV jitter. This finding is in agreement with the previous work of \citet{Bastien-14} and \citet{Cegla-14}.

Moreover, we examined the capability of two state-of-the-art modeling techniques, namely the FF' method and \textit{SOAP2.0}, in accurately predicting the RV jitter amplitude using simultaneous photometric observations. We conclude that for very active stars both techniques can reasonably well predict the amplitude of RV jitter.  Furthemore, we find that the \textit{FF'} method systematically underpredicts the RV jitter amplitude across the whole range of photometric variability levels considered, and this is particularly apparent for the low activity stars in our sample. This finding is in strong agreement with the \citet{Bastien-14} result. Furthermore, the accuracy of the \textit{SOAP2.0} predictions assuming a single spot model improves significantly when the photometric
variability rises above 1 ppt limit. On the contrary for quiet stars, when the photometric variability is smaller than 1 ppt, the \textit{SOAP2.0} single plage model provides more accurate predictions. Thus, our results provide stronger evidence (and support of) the existence of an activity level boundary where spots are spot-dominated at higher activity levels, transitioning to plage-dominated at lower activity levels.

\begin{acknowledgements}

\scriptsize MO acknowledges research funding from the Deutsche
Forschungsgemeinschft (DFG , German Research Foundation) - OS 508/1-1. This work was supported by  Funda\c{c}\~ao para a Ci\^encia e a Tecnologia (FCT, Portugal) through the research grant through national
funds and by FEDER through COMPETE2020 by grants UID/FIS/04434/2013
\& POCI-01-0145-FEDER-007672, PTDC/FIS-AST/1526/2014 \& POCI-01-
0145-FEDER-016886 and PTDC/FIS-AST/7073/2014 \& POCI-01-0145-
FEDER-016880. P.F., N.C.S., V.A. and S.B. acknowledge support from FCT
through Investigador FCT contracts nr. IF/01037/2013CP1191/CT0001, IF/00169/2012/CP0150/CT0002, IF/00650/2015/CP1273/CT0001, and IF/01312/2014/CP1215/CT0004. PF further acknowledges
support from Funda\c{c}\~ao para a Ci\^encia e a Tecnologia (FCT) in the
form of an exploratory project of reference IF/01037/2013CP1191/CT0001. P.F.  acknowledges support from  FCT transnational cooperation program Pessoa. M.O  also acknowledges the support
of COST Action TD1308 through STSM grant with reference Number:STSM-TD1308-030416-077992. HMC acknowledges the financial support of the National Centre for Competence in Research “PlanetS” supported by the Swiss National Science Foundation (SNSF). CAW acknowledges support from the STFC grant ST/P000312/1. We would like to
thank the anonymous referee for insightful
suggestions, which added significantly to the clarity of
this paper. Last but not least, we would like to thank Fabienne A. Bastien for kindly providing the F8 measurements of stars in \citet{Bastien-14}.

\end{acknowledgements}

\bibliographystyle{aa}
\bibliography{mahlibspot}

\begin{appendix}
\section{Correlation and phase shift}
In this appendix we aim to probe the dependency of the correlation between two observables on the phase shift between them. To assess this we generated two identical sinusoidal signals and examined the variation of the Spearman rank-order correlation coefficient by introducing a phase-shift between the two signals. Figure A.1 demonstrates how the Spearman's correlation coefficient varies as a function of phase shift, and goes from a perfect correlation to anti-correlation in half a phase shift. \citet{Queloz-09} estimated a one-quarter of a stellar rotation phase shift between RVs and FWHM, therefore, in Figure A.1 we present that region with a green shaded area. As one can see, even a quarter of a stellar rotation phase difference between observables and RVs is sufficient to  extinguish the strong correlation that might exist between them. However, we would like to note that activity signals are not typically sinusoidal, and therefore even if they were not shifted the correlation between RV and any indicator are usually not correlated
with a 1:1 relation and the correlation coefficient would never be 1.

\begin{figure}[!h]
\includegraphics[width=0.45\textwidth, height=60mm]{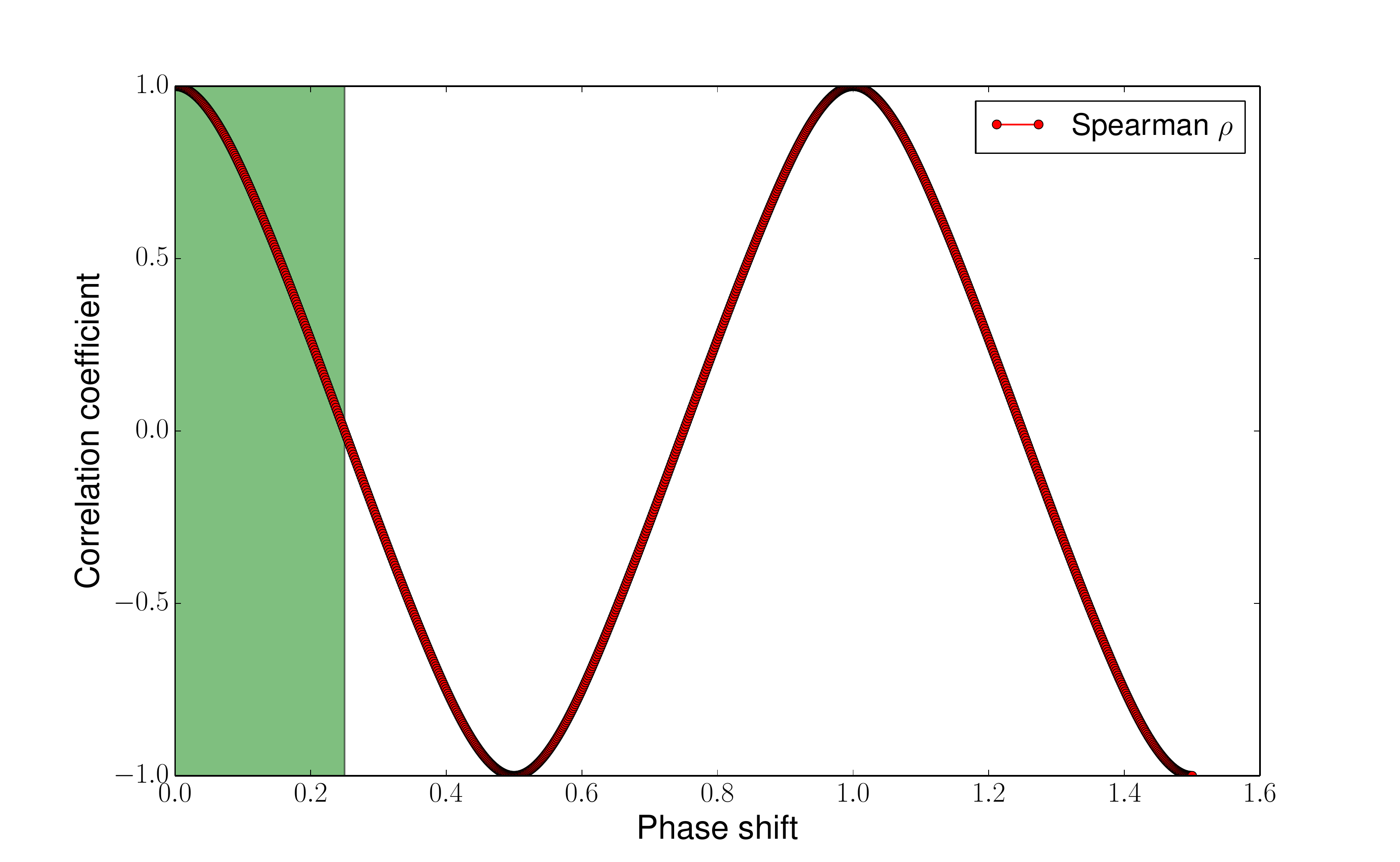}

 \caption{The variation of the Spearman's correlation coefficient as a function of the phase shift between two identical sinusoidal signals. The green shaded area represents the one-quarter of phase shift.}
  \label{sample-figure}
\end{figure}

\end{appendix}

\end{document}